\documentclass[onecolumn]{emulateapj}
\usepackage{amsfonts}
\usepackage{amsmath}
\usepackage{amssymb}
\usepackage{wrapfig}
\usepackage{graphicx}
\usepackage{bibentry,natbib}
\usepackage{wasysym}
\usepackage{latexsym}
\usepackage{subfigure}

\newcommand{\half}{{\textstyle{1\over2}}}

\newcommand{\ms}{\noalign{\vspace{3pt plus2pt minus1pt}}}

\newcommand{\be}{\begin{equation}}
\newcommand{\ee}{\end{equation}}
\newcommand{\bea}{\begin{eqnarray}}
\newcommand{\eea}{\end{eqnarray}}
\newcommand{\nn}{\nonumber}

\def\bi{\bf}
\def\grad{{\rm grad}\,}
\def\div{{\rm div}\,}
\def\curl{{\rm curl}\,}

\slugcomment{to be submitted to Astrophys. J.}

\shorttitle{Inductive electric field}
\shortauthors{Melrose}

\begin{document}

\title{Magnetic explosions: role of the inductive electric field}

\author{D. B.  Melrose}

\affil{Sydney Institute for Astronomy, School of Physics, The University of Sydney, NSW 2006, Australia}

\begin{abstract}
Inclusion of the inductive electric field, ${\bf E}_{\rm ind}$ due to the temporally changing ${\bf B}$, in magnetic explosions is discussed, with emphasis on solar flares. Several roles played by ${\bf E}_{\rm ind}$ are identified:  on a global scale, ${\bf E}_{\rm ind}$ produces the EMF that drives the explosion;  the associated ${\bf E}_{\rm ind}\times{\bf B}$ drift is identified with the inflow of magnetic field lines into a reconnection region;  the polarization current, associated with $\partial{\bf E}_{\rm ind}/\partial t$, implies a ${\bf J}\times{\bf B}$ force that accelerates this inflow; and the component of ${\bf E}_{\rm ind}$ parallel to ${\bf B}$ accelerates the energetic electrons that cause hard X-ray emission and type~III radio bursts. Some simple models that describe these effects are presented. A resolution of the long-standing ``number problem'' in solar flares is suggested.
\end{abstract}


%%%%%%%%%%%%%%%%%%%%%%%%%%%%%%%%%%%%%%%%%%%%%%%%%%%

\section{Introduction}
\label{section:introduction}

As discussed in an accompanying paper, referred to as paper~1, magnetic explosions are intrinsically time-dependent phenomena, and must involve an inductive electric field, ${\bf E}_{\rm ind}$, and a displacement current ${\bf J}_{\rm disp}=\varepsilon_0\partial{\bf E}_{\rm ind}\partial t$. These fields are not taken into account explicitly in conventional models for solar flares and other magnetic explosions; the physics associated with them is either treated in terms of quasi-electrostatic proxies or ignored. The following six roles for ${\bi E}_{\rm ind}$ and ${\bf J}_{\rm disp}$ need to be taken into account in models for magnetic explosions. 
\begin{enumerate}
\item On a global scale, the electromotive force (EMF), $\Phi=-d\Psi_B/dt$, that drives the explosion-related current is determined by the rate of change of the magnetic flux, $\Psi_B$, enclosed by the circuit around which this current flows.

\item The electric drift, ${\bi u}_{\rm ind}={\bi E}_{\rm ind}\times{\bf B}/|{\bf B}|^2$, associated with the inductive field transports magnetic energy into the energy-release (or reconnection) region: no mechanical driver is required.

\item The collective response of the plasma particles to the displacement current results in the polarization current, ${\bf J}_{\rm pol}=(c^2/v_A^2){\bf J}_{\rm disp\perp}$, where $v_A$ is the Alfv\'en speed.

\item The ${\bf J}\times{\bf B}$ force due to the polarization current accelerates ${\bi u}_{\rm ind}$, providing the link that allows the changing magnetic field to drive the magnetic energy release.

\item The Alfv\'enic energy flux that transports energy from the energy-release region to the acceleration region is set up by the polarization current changing the current profile in the linking flux tube, and driving a vortex motion of the plasma within the flux tube.

\item Acceleration by $E_\parallel$ relies on ${\bi E}_{\rm ind}$ and ${\bf J}_{\rm disp}$ \citep{SL06}. 
\end{enumerate}
The first of these roles is discussed in paper~1, and the other five are discussed in this paper. 

A long-standing ``number problem'' concerns the acceleration of the electrons that produce hard X-rays \citep{MB89}: the number of accelerated electrons inferred from X-ray data exceeds the number available by a large factor, $M=10^6$ assumed here. A way of resolving the number problem, by including a return current in the model for Alfv\'enic energy transport, is suggested in \S\ref{section:acceleration}.

The argument for the inductive field being the flare driver is presented in \S\ref{section:driver}.  The role of the polarization current in driving this inflow is discussed in \S\ref{section:energyflow}.  The role of time-dependent fields in Alfv\'enic energy transport to the acceleration region is discussed in \S\ref{section:transport}. Acceleration of electrons by $E_\parallel$, including the number problem, is discussed in \S\ref{section:acceleration}. The results are discussed and summarized in \S\ref{section:conclusions}.

\section{Inductive field as a flare driver}
\label{section:driver}
In this section it is assumed that the parallel component of the inductive field is screened by charges. The perpendicular component is interpreted in terms of the ${\bf E}\times{\bf B}$ drift velocity, ${\bi u}_{\rm ind}$. It is argued that ${\bi u}_{\rm ind}$ can be identified as the inflow velocity, caused by the time-changing magnetic field, into the reconnection region. This contrasts with a conventional steady-state model is which the inflow is imposed as a boundary condition, and is attributed to a mechanical driver.

\subsection{Inductive drift}

The assumption that the parallel component of the inductive electric field is screened implies the presence of a charge density, $\rho$. This charge density is such that it produces a parallel electric field that is equal and opposite to the parallel component, ${\bi E}_{{\rm ind}\parallel}$, of the unscreened ${\bf E}_{\rm ind}$. The remaining (perpendicular) component of the inductively-induced electric field has a nonzero divergence
\be
\curl{\bi E}_{{\rm ind}\,\perp}=-{\partial{\bi B}\over\partial t},
\qquad
\div{\bi E}_{{\rm ind}\,\perp}={\rho\over\varepsilon_0}.
\label{if1}
\ee
The inductive drift is the electric drift due to ${\bi E}_{{\rm ind}\,\perp}$:
\be
{\bi E}_{{\rm ind}\,\perp}=-{\bi u}_{{\rm ind}}\times{\bi B},
\qquad
{\bi u}_{{\rm ind}}={{\bi E}_{{\rm ind}\,\perp}\times{\bi B}\over|{\bi B}|^2}.
\label{if2}
\ee
Combining (\ref{if1}) and (\ref{if2}) gives
\be
{\partial{\bi B}\over\partial t}= \curl({\bi u}_{{\rm ind}}\times{\bi B}),
\label{if3}
\ee
which is the equation for magnetic field lines moving with the velocity ${\bi u}_{{\rm ind}}$. The screening charge density is
\be
\rho=\varepsilon_0[{\bi u}_{{\rm ind}}\cdot\curl{\bi B}-{\bi B}\cdot\curl{\bi u}_{{\rm ind}}].
\label{if4}
\ee

\begin{figure} [t]
\centerline{
\includegraphics[width=7cm]{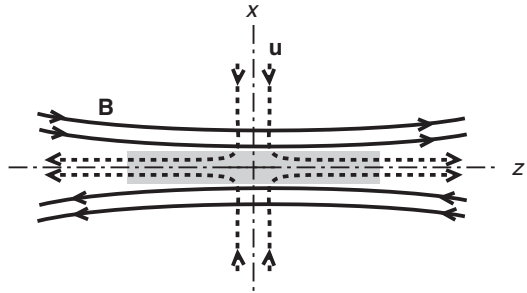}
}
\caption{A cartoon of the Sweet-Parker reconnection model: the magnetic field (solid) and flow (dashed) lines are confined to the $x$-$z$ plane; dissipation is confined to a current sheet, shown in gray, about the magnetic neutral plane, $x=0$.}
\label{fig:Sweet-Parker}
\end{figure} 

\subsection{Sweet-Parker-type model}

Consider any steady-state reconnection model, such as the Sweet-Parker-type model illustrated in Fig.~\ref{fig:Sweet-Parker}. Such a model involves an inflow of plasma and magnetic field into a current sheet, with this inflow being specified as a boundary condition. Physically, the inflow must be driven in order for fast reconnection to proceed, and in a steady-state model this requires a mechanical driver acting across field lines, usually assumed to be a pressure gradient \citep{S78,H94}. However, the pressure gradient is much smaller that the available Maxwell stress, due to the ${\bi J}\times {\bi B}$ force, and the latter is a more plausible driver, as proposed in Uchida's quadrupolar model \citep{U96,Hetal01} for an erupting filament. Here it is suggested that the ${\bi J}\times {\bi B}$ force acts as a driver under more general conditions, provided that the (artificial) steady-state assumption is abandoned.

A simple model that illustrates the role of ${\bi u}_{{\rm ind}}$ plays in reconnection involves a magnetic field, that varies globally as a function of time, $f(t)$. Although such an assumption is also artificial, it complements a steady-state model. The function $g(t)=df(t)/dt$ can be regarded as simulating the rate of reduction in stored magnetic energy beyond the boundary of the steady-state region; the associated ${\bi u}_{{\rm ind}}$ becomes the inflow velocity.

In the simplest example of a planar geometry, let the magnetic field along the $z$ axis, and oppositely directed above and below the current sheet at $x=0$. Assuming that the $x$- and $t$-dependences are independent, one may write $B_z(x,t)=f(t)b(x)$, with $b(-x)=-b(x)$. Assuming that the system is in uniform in $y$, one has
\be
E_{{\rm ind}\,y}(x,t)=-g(t)\int_0^xdx'\,b(x'),
\qquad
g(t)={df(t)\over dt},
\label{if5}
\ee
The inductive drift,
\be
u_{{\rm ind}\,x}(x,t)={g(t)\over f(t)b(x)}\,\int_0^xdx'\,b(x'),
\label{if6}
\ee
is along the $x$-axis, with $u_{{\rm ind}\,x}(-x,t)=-u_{{\rm ind}\,x}(x,t)$. For a decreasing field, $g(t)<0$, the flow is directed towards the current sheet at $x=0$. A generalization to allow the magnetic field lines to be curved in the $x$-$z$ plane is needed to allow outflow along the $z$ axis. Assuming $B_x(x,z,t)=f(t)b_x(x,z)$, $B_z(x,z,t)=f(t)b_z(x,z)$, the inductive electric field is again along the $y$ axis, 
\be
E_{{\rm ind}\,y}(x,z,t)=-g(t)\left[\int_0^xdx'\,b_z(x',z)
-\int_0^zdx'\,b_x(x,z')\right],
\label{if7}
\ee
and ${\bi u}_{{\rm ind}}$ has a component along the $z$-axis, such that ${\bi u}_{{\rm ind}}\cdot{\bi B}=0$. In both these simple models there is no parallel component of the inductive electric field,  so that screening of the parallel component of the inductive field is not relevant, and one has $\rho=0$.

\subsection{Self-similar model}

A generalization of the foregoing discussion is possible in the special case where the change in the magnetic field is self-similar. Specifically, a self-similar model corresponds to ${\bi B}({\bi x},t)=f(t){\bi b}({\bi x})$. The inductive electric field can then be written ${\bi E}_{{\rm ind}}({\bi x},t)=g(t){\bi e}({\bi x})$. The inductive drift is 
\be
{\bi u}_{{\rm ind}}({\bi x},t)={g(t)\over f(t)}\,{{\bi e}({\bi x})\times{\bi b}({\bi x})\over|{\bi b}({\bi x})|^2}.
\label{ssm1}
\ee
The assumed screening of the parallel component of ${\bi E}_{{\rm ind}}$ implies  ${\bi e}({\bi x})\cdot{\bi b}({\bi x})=0$ and requires $\rho({\bi x},t)=g(t)r({\bi x})\ne0$. Introducing the vector potential, ${\bi A}({\bi x},t)=g(t){\bi a}({\bi x})$, and the scalar potential, $\phi({\bi x},t)=g(t)p({\bi x})$ in the Coulomb gauge, $\div{\bi a}({\bi x})=0$, one requires that $p({\bi x})$ depend on the distance, $s$, along field lines, with $\partial^2p({\bi x})/\partial s^2=r({\bi x})/\varepsilon_0$. One then has
\be
{\bi e}({\bi x})=-{\bi a}({\bi x}),
\qquad
{\bi b}({\bi x})=\curl{\bi a}({\bi x}).
\label{ssm2}
\ee
The parallel-screening condition becomes
\be
{\bi a}({\bi x})\cdot\curl{\bi a}({\bi x})=0.
\label{ssm3}
\ee
The flow (\ref{ssm1}) is along the direction of
\be
{\bi a}({\bi x})\times\curl{\bi a}({\bi x})=\grad(\half|{\bi a}({\bi x})|^2)
-[{\bi a}({\bi x})\cdot{\rm grad}]\,{\bi a}({\bi x}).
\label{ssm4}
\ee

The condition (\ref{ssm3}) restricts the form of a magnetic field that can evolve in a self-similar manner. Seeking a solution that is separable in Cartesian coordinates, one finds that (\ref{ssm3}) is satisfied by
\bea
{\bi a}({\bi x})&=&(YZ,XZ,XY),
\nn
\\
\ms
{\bi b}({\bi x})&=&(X[Y'-Z'],Y[Z'-X'],Z[X'-Y']),
\label{ssm5}
\eea
where $X=X(x)$, $Y=Y(y)$, $Z=Z(z)$ are arbitrary functions, and $X',Y',Z'$ are their derivatives. The model corresponding to (\ref{if7}) is reproduced by $Y=0$, and the model corresponding to (\ref{if5}) is reproduced by $Y=0$, $Z=1$.

\section{Role of the polarization current}
\label{section:energyflow}

In this section, the role of the polarization current associated with a changing inductive electric field is discussed. 

\subsection{Polarization current}

The presence of a time-dependent electric field in a plasma causes a charged particle, with charge $q$ and mass $m$, to drift across the magnetic field. This polarization drift is \citep{N63}
\be
{\bf v}_{\rm pol}={m\over q|{\bf B}|^2}{d{\bi E}_\perp\over d t}.
\label{vp}
\ee
In the following, $d{\bi E}_\perp/d t$ is identified as $\partial{\bi E}_{\rm ind\perp}/\partial t$, whose curl is determined by $\partial^2{\bf B}/\partial t^2$. The drift (\ref{vp}) depends on the sign of the charge, and summing over all charges it implies a current that is proportional to the displacement current. This current is referred to as the polarization current:
\be
{\bi J}_{{\rm pol}}={c^2\over v_A^2}\,\varepsilon_0{\partial{\bi E}_\perp\over\partial t}.
\label{Jp}
\ee

This polarization current may be interpreted as the response of the plasma to the displacement current, and an alternative derivation of (\ref{Jp}) follows by considering the low-frequency response of a plasma. The susceptibility tensor for a magnetized plasma at very low frequency is approximately diagonal, with components $c^2/v_A^2,c^2/v_A^2,-\omega_p^2/\omega^2$, where $\omega_p$ is the plasma frequency. This tensor relates the polarization, ${\bf P}/\varepsilon_0$, to the electric field, with the current density related to the polarization by ${\bf J}=\partial{\bf P}/\partial t$. Differentiating the perpendicular components with respect to time, assuming for simplicity that $v_A^2$ is a constant, and integrating the parallel component (with $-\omega^2\to\partial^2/\partial t^2$) implies that the response can be rewritten in the form
\be
{\bi J}_\perp={c^2\over v_A^2}\,\varepsilon_0{\partial{\bi E}_\perp\over \partial t},
\qquad
{\partial J_\parallel\over\partial t}=\varepsilon_0\omega_p^2E_\parallel.
\label{pol1}
\ee
This provides an alternative  derivation of ${\bf J}_{\rm pol}={\bf J}_\perp$. This alternative formalism allows one to include dissipation, through the conductivity tensor, with perpendicular component $\sigma_\perp$ and parallel component $\sigma_\parallel$, leading to additional terms, $\sigma_\perp{\bf E}_\perp$ and $\sigma_\parallel \,\partial E_\parallel/\partial t$, on the right hand sides of the two equations (\ref{pol1}), respectively.

On differentiating the expression (\ref{if2}) for ${\bf u}_{\rm ind}$ with respect to time, neglecting the time-dependence of ${\bf B}$, the result becomes
\be
\eta{d{\bf u}_{\rm ind}\over dt}={\bf J}_{\rm pol}\times{\bf B},
\label{pol2}
\ee
where $\eta$ is the mass density. It follows that the polarization current provides the force that accelerates ${\bf u}_{\rm ind}$. 

These results lead to the following interpretation of the energy inflow into the energy-release region. Magnetic energy is stored in a global current system, and some trigger causes ${\bi B}$ to start decreasing in time. The associated ${\bf J}_{\rm pol}$ accelerates an inductive flow ${\bi u}_{\rm ind}$ that transports energy over large distances to the (ill-defined) boundary of the energy-release region. Within the energy-release region the energy flow may be described by a steady-state model, with further acceleration by ${\bf J}_{\rm pol}$ being unimportant. The rate of decrease in the globally stored energy is balanced by the outflow of energy from the energy-release region. 

The conceptual change from a conventional model is that no pressure gradient or other mechanical force is need to drive the inflow. As the energy release begins, ${\bf J}_{\rm pol}$ becomes nonzero, and the inflow is driven by the ${\bf J}_{\rm pol}\times{\bi B}$ force. 

\subsection{Energy-release region}
\label{sect:energy-release}

The conversion of magnetic energy into the energy in nonthermal particles in a magnetic explosion is assumed to occur in two stages: conversion of magnetic energy into an Alfv\'enic flux in an energy-release region, and conversion of the Alfv\'enic flux into energetic particles in a remotely located acceleration region. The energy conversion in the first of these region is discussed here.

Conversion of magnetic energy into another form requires some dissipation, but the dissipation need not be energetically significant. Reconnection is effective only around a magnetic null \citep{PF00}, and  the Sweet-Parker model illustrated in Fig.~\ref{fig:Sweet-Parker} shows the inflow and outflow in such a region. To be effective in a solar flare, reconnection must be fast (compared with the Sweet-Parker rate), and the geometry is similar for fast reconnection models. Numerical simulations show that  fast (Petschek) reconnection can occur only if the resistivity is  localized \citep{BS01,U03} and anomalous.\citep{K01,K05}, in which case the reconnection region is on a micro scale (paper~1).  A statistically large number of such reconnection sites is needed, and this requires multiple local magnetic nulls \citep{LV99}. Although dissipation is an essential ingredient in reconnection, the energy dissipated in converting the magnetic energy into kinetic energy or an Alfv\'enic flux  is arbitrarily small for arbitrarily small regions of anomalous resistivity.  It is assumed here that  magnetic energy conversion occurs without significant dissipation in the energy-release region.

The energy flow into the energy-release region is inductively driven, as discussed above. Within the energy-release region a steady-state model relates the energy outflow to the inflow. The outflow can be in three forms: kinetic energy, magnetic energy transported due to field lines frozen into the flow, and an Aflv\'enic flux. In conventional flare models, such as the CSHKP model,  the downward moving newly-closed field lines ultimately  join the underlying closed magnetic structure. This requires that they slow down, reducing the kinetic energy in the flow. The length of the reconnected flux tubes also decreases, and the magnetic energy associated with the current flowing along these field lines also reduces as they are dragged into their final location. This implies that both the kinetic energy initially in the outflow, and the magnetic energy transported by the outflow decrease. It is assumed here that the definition of the energy-release region is extended to include this further energy conversion. It follows that the relevant energy emerges as an Alfv\'enic flux, propagating along the newly closed field lines. These remarks support earlier suggestions that the outflowing energy is Alfv\'enic \citep{ES82,H94,FH08,KHB11}; however, there is no detailed model that includes all these effects. A specific model for the Alfv\'enic energy transport away from the energy-release region is presented in the next section.

\section{Alfv\'enic energy transport}
\label{section:transport}

In this section, a specific (coaxial) model is developed for the Alfv\'enic transport of energy from the energy-release region to the acceleration region. Although highly idealized, this model shows how the large power released can be transported over large distances.

\subsection{Setting up the Alfv\'enic flux}

The magnetic energy inflow into the energy-release region is balanced by an Alfv\'enic outflow. A complication in developing simple models for the inflow and outflow is geometric: the inflow into the energy-release region is modeled using cartesian coordinates, and far from the energy-release region the Alfv\'enic flow is more appropriately modeled using cylindrical coordinates. Rather than attempt to formulate a more general geometry that covers both regions,  cartesian and cylindrical geometries are simply assumed for the inflow and outflow, respectively.

Consider a cylindrical model for a flux tube of radius $R$ with radial coordinate $r<R$, azimuthal angle $\phi$ and axial distance $z$ \citep{M92}. The magnetic field is twisted due to the current, and it has components $B_\phi,B_z$. For simplicity the twist is assumed to be small, $|B_\phi|\ll|B_z|$. The pre-flare current is assumed to be force-free, implying $J_\phi/J_z=B_\phi/B_z$, and the assumption that the twist is small allows one to calculate $B_\phi$ in terms of $J_z$ and ignore $J_\phi$. It is convenient to write the radial dependence of the current density in the form $J_z(r)=J_0j_1(\xi)$, with $\xi=r/R$ a dimensionless radial distance. The function $j_1(\xi)$ describes the current profile in the flux tube prior to the flare. Prior to the flare, there is no electric field, no plasma motion and no energy flux. At the onset of the flare, the inductive electric field is set up, and this is identified as a radial field, $E_r\propto g(t)$. The displacement current, $\varepsilon_0\partial E_r/\partial t\propto dg(t)/dt$, drives a radial polarization current, $J_r\propto dg(t)/dt$. This radial current transfers current across field lines, changing the current profile as a function of time, concentrating the current density at smaller radial distances, thereby increasing $B_\phi$ at $r<R$. 

The radial electric field implies a vortex motion, $u_\phi\approx-E_r/B_z\propto g(t)$, where $B_z$ is assumed to be independent of time within the cylinder. During the onset of the energy release, $u_\phi$ increases in magnitude, driven by the force $-J_r B_z\propto dg(t)/dt$. The vortex motion increases the twist, $B_\phi/B_z$. Thus the onset of the energy release launches a propagating twist, which may be interpreted as an Alfv\'enic front that sets up the Alfv\'enic flux. The front and the subsequent energy flux propagate at the MHD speed, $v_0=v_A/(1+v_A^2/c^2)^{1/2}$. The profile of the front may be described by a function $g(t-z/v_0)$,  whose value at $z=0$ is $df(t)/dt$.  After the initial transient phase, described by $dg(t)/dt=d^2f(t)/dt^2$, it is assumed that the energy flux approaches a constant, with $g(t)\to1$. Before discussing the Alfv\'enic flux further, it is appropriate to introduce a specific model for the change in the current profile.

\begin{figure} [t]
\centerline{
\includegraphics[scale=0.6]{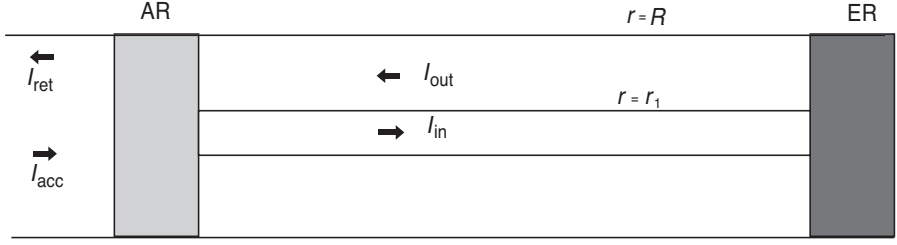}
}
\caption{A schematic of the coaxial model for Alfv\'enic energy transport in the upward current region ($I_0>0$). The dark box  on the right is the energy-release (ER) region, the central section is the coaxial model, with a current $I_{\rm in}$ at $r<r_1$, and a current $I_{\rm out}=I_0-I_{\rm in}$ (indicated as a return current) at $r_1<r<R$; the lightly shaded box on the left is the acceleration region (AR), and on the far left is the current, $I_{\rm acc}$ carried by precipitating electrons, and the return current, $I_{\rm ret}=I_0-I_{\rm acc}$ from the photosphere. }
\label{fig:coaxial}
\end{figure} 

\subsection{Change in axial current profile}

Let the current profile when the Alfv\'enic flux is established ($g(t)\to1$) be described by $j_2(\xi)$. As the current profile changes at $z=0$, it may be described by $j_1(\xi)[1-g(t)]+j_2(\xi)g(t)$, with $g(t)=0$ prior to the onset of the energy release, and with $g(t)\to1$ after the initial transient phase. At points $z>0$ the changing current profile is described by
\be
J_z=J_0\{j_1(\xi)[1-g(t-z/v_0)]+j_2(\xi)g(t-z/v_0)\}.
\label{Af0}
\ee
The axial component of the Poynting vector, which describes the energy flux, is $S_z=E_r B_\phi/\mu_0$. The azimuthal component of the magnetic field, before and after the Alfv\'enic flux is set up, is 
\be
[B_\phi(r)]_{1,2}={\mu_0I_0\over\pi R}b_{1,2}(\xi),
\qquad
b_{1,2}(\xi)={1\over \xi}\int_0^\xi d\xi'\xi'j_{1,2}(\xi'),
\label{Af1}
\ee
with $I_0=\pi R^2J_0$. The radial electric field is related to $B_\phi$ through $\partial E_r/\partial z= \partial B_\phi/\partial t$, and assuming  $E_r\propto g(t-z/v_0)$, this gives
\be
E_r=v_0(B_{\phi2}-B_{\phi1})g,
\label{Af2}
\ee
with $g=g(t-z/v_0)$ and $B_{\phi1,2}=[B_\phi(r)]_{1,2}$. The Poynting flux is then
\be
S_z={v_0\over\mu_0}[B_{\phi2}-B_{\phi1}][gB_{\phi2}+(1-g)B_{\phi1}]g,
\label{Af3}
\ee
After an initial transient phase, the Alfv\'enic flux approaches a constant value $v_0[B_{\phi2}-B_{\phi1}]B_{\phi2}/\mu_0$.

\subsection{Coaxial model}
\label{sect:coaxial}

A simple model for the current profile has $j_1(\xi)$ independent of $\xi$, and $j_2(\xi)$ equal to one constant for $\xi<\xi_1$ and another constant for $\xi_1<\xi<1$, as illustrated in Fig.~\ref{fig:coaxial}. In fact, an arbitrary force-free cylindrical current profile can be approximated by a set of concentric rings with the current density a constant within each ring, with solution for $B_\phi$  involving Bessel functions of the first and second kind \citep{MNB94}. The simple model proposed here is a reasonable approximation to such a model for a current profile that is strongly concentrated in an inner region, $r<R_1=R\xi_1$, provided that the twist is sufficiently small, that is, for $B_\phi/B_z\ll1$. It is convenient to introduce the current, $I_{\rm in}$, in the inner region, $r<R_1$, and the current, $I_{\rm out}$, in the outer region, $R_1<r<R$, with $I_{\rm in}+I_{\rm out}=I_0$, where $I_0$ is the current in the flux tube prior to the onset of the flare. 

After the onset phase, as $g(t)$ approaches unity, the current density is assumed to approach the constant values $aJ_0$ for $\xi<\xi_1$ and $bJ_0$ for $\xi_1<\xi<1$. This model involves three parameters, $a,b,\xi_1$, only two of which are independent, with $(a-b)\xi_1^2=1-b$. For $g(t)\to1$, this model gives, in terms of (\ref{Af1}),
\be
b_2(\xi)=\left\{
\begin{array}{ll}
a\xi/2,&\;\xi<\xi_1,
\\
\ms
b\xi/2+(1-b)/2\xi,&\,\xi_1<\xi<1.
\end{array}
\right.
\label{Af4}
\ee
With $b_1(\xi)=\xi/2$, (\ref{Af2}) with (\ref{Af1}) gives
\be
E_r=-{\mu_0v_0I_0\over\pi R}\left\{
\begin{array}{ll}
(a-1)\xi/2,&\;\xi<\xi_1,
\\
\ms
(1-b)(1-\xi^2)/2\xi,&\,\xi_1<\xi<1.
\end{array}
\right.
\label{Af5}
\ee
Introducing the potential $\Phi(r)$, such that $E_r=-d\Phi(r)/dr$, (\ref{Af5}) with $\Phi(0)=0$ implies
\be
\Phi(r)={\mu_0v_0I_0\over4\pi}\left\{
\begin{array}{ll}
(a-1)\xi^2,&\;\xi<\xi_1,
\\
\ms
(1-b)[\ln(\xi^2/\xi_1^2)+1-\xi^2],&\;\xi_1<\xi<1.
\end{array}
\right.
\label{Af6}
\ee
The power at $r<\xi R$ in the Alfv\'enic flux is
\bea
&&2\pi R^2\int_0^\xi d\xi'\xi'S_z(\xi')={\mu_0v_0I_0^2\over4\pi}(1-b)^2s(\xi),
\nn
\\
\ms
&&s(\xi)=\left\{
\begin{array}{ll}
(a-1)\xi^4/2(1-b)^2,&\;\xi<\xi_1,
\\
\ms
\displaystyle{
\ln{\xi^2\over \xi_1^2}
-{(1-2b)(\xi^2-\xi_1^2)
\over1-b}- {b(\xi^4-\xi_1^4)\over2(1-b)}},&\;\xi_1<\xi<1.
\end{array}
\right.
\label{Af7}
\eea

The model involves two independent parameters, $b$ and $\xi_1$, with $a$ related to them by $(a-b)\xi_1^2=1-b$. There is zero Alfv\'enic flux for $b=1$, corresponding to the initial current profile; the power increases with decreasing $b$, as the current becomes more concentrated near the axis. For $b=0$, corresponding to all the current in the inner region $r<\xi_1R$, the power in the Alfv\'enic flux is of order $I_0\Phi_0$ with $\Phi_0=\Sigma_AI_0$, which is consistent with the total power released in a flare. In this way, the model shows how the released magnetic energy can be transported along field lines in a quasi-stationary manner. The parameter $b$ can be negative, as discussed in the next section in connection with a return current and the number problem. 

The parameter $\xi_1$ determines the inner radius, $\xi_1R$, and is limited by two requirements. First, the drift speed of the electrons, $aJ_0/en_e$, relative to the ions must not exceed the threshold for the development of an instability that leads to the onset of anomalous resistivity. Second, the twist $B_\phi/B_z$ must not exceed the threshold (of order unity) for the current-carrying region to become kink unstable. 

\subsection{Inclusion of the displacement current}

The displacement current is usually ignored in MHD, and it is important to include it in discussing acceleration by $E_\parallel$. This was emphasized by \cite{SL06}, who derived two coupled equations for the $t$ and $z$ derivatives of the parallel components of the net current density and the vorticity, ${\bf\Omega}=\curl{\bf u}$. In the cylindrical model assumed above, the parallel vorticity is $\Omega_z=r^{-1}\partial[r u_\phi(r)]/\partial r$.

Neglecting terms that are not important in the present discussion, the equations derived by \cite{SL06} are
\be
{\partial J'_\parallel\over\partial t}\approx{B_z\over\mu_0}{\partial\Omega_z\over\partial z},
\qquad
{\bf J}'={\bf J}+\varepsilon_0{\partial{\bf E}\over\partial t},
\label{SL1}
\ee
where the net current ${\bf J}'$ includes the displacement current, and with
\be
\left(1+{c^2\over v_A^2}\right){\partial\Omega_z\over\partial t}\approx{\mu_0c^2\over B_z}{\partial J'_\parallel\over\partial z}.
\label{SL2}
\ee
On differentiating either (\ref{SL1}) or (\ref{SL2}) with respect to $t$ and using the other, one finds that $J'_\parallel$ and $\Omega_z$ satisfy the equation for an Alfv\'en wave propagating at $v_0=v_A/(1+v_A^2/c^2)^{1/2}$. This justifies the assumption made above that the temporal variation $g(t)$ at $z=0$ propagates away as $g(t-z/v_0)$ at $z>0$.

\begin{figure} [t]
\centerline{
\includegraphics[scale=0.3]{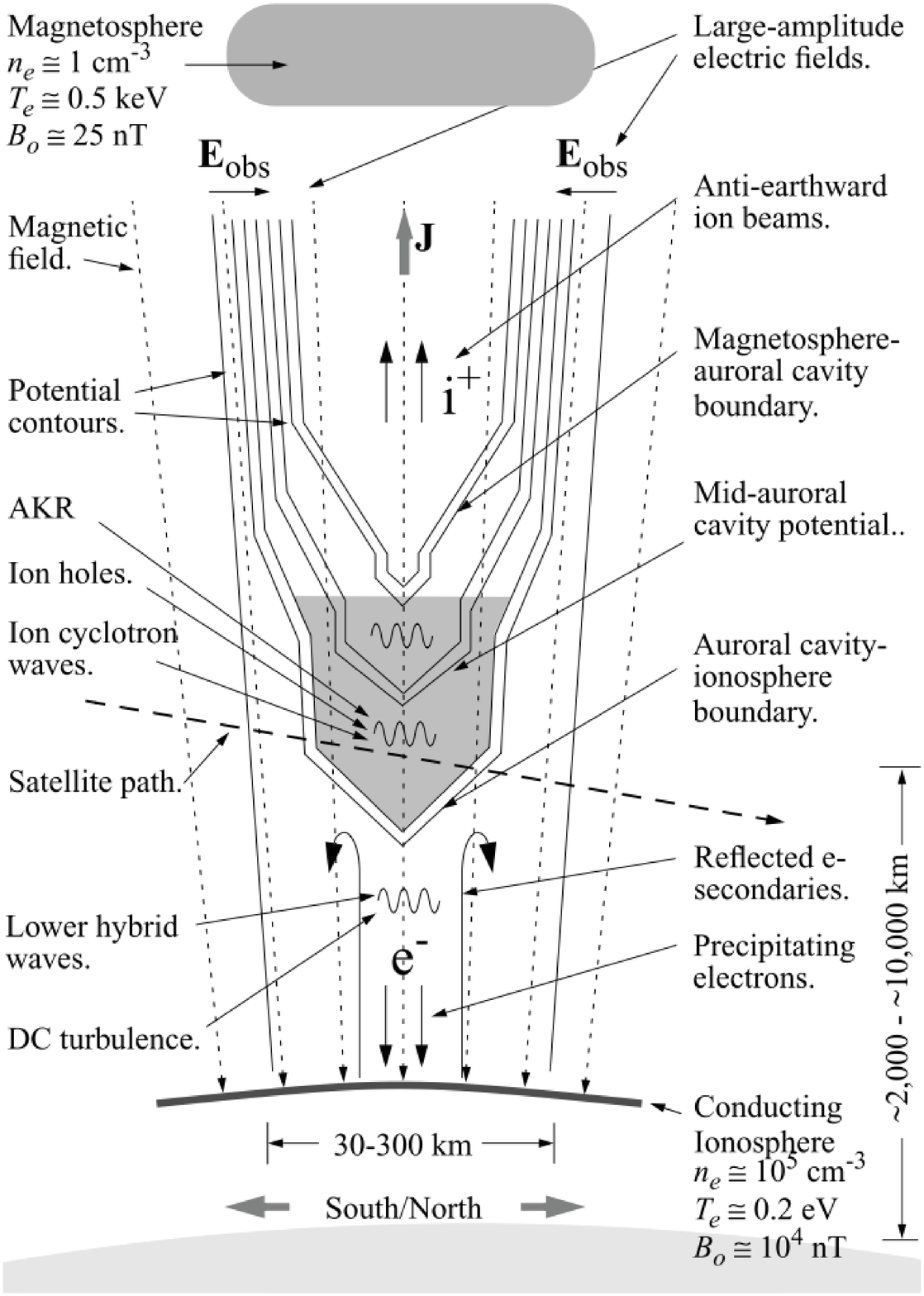}
}
\caption{A cartoon of the upward current region of the aurora. The potential contours (solid lines) indicate a low-altitude and high-altitude acceleration region with the auroral cavity in between. [After \cite{Eetal04}]}
\label{fig:auroral}
\end{figure} 

\section{Acceleration/dissipation region}
\label{section:acceleration}

The acceleration of electrons in a solar flare is an outstanding unsolved problem. In this section, the ingredients needed to solve this problem within the framework of the generic model are discussed. The acceleration mechanism is assumed to be analogous to the mechanism for auroral electrons, due to $E_\parallel$, in a region remote from the energy-release region.

\subsection{Requirements on the acceleration region}

The following macroscopic requirements are plausible for the acceleration region.
\begin{enumerate}
\item The Alfv\'enic flux, incident on top of the region decreases to zero at the bottom of the region. 
\item The potential is across field lines at the top of the region, is zero at the bottom of the region, and is partially along field lines within the region, producing $E_\parallel$. 
\item Electrons are accelerated by $E_\parallel$  to $\varepsilon=e\Delta\Phi$, $\Delta\Phi=\Phi/M$, with $M\gg1$.
\item The available power, $I_0\Phi_0$, is transferred to electrons, accelerated at the rate ${\dot N}=MI_0/e$.
 \end{enumerate}
 The most challenging of these constraints is the large value of $M$, which may be interpreted in terms of the ``number problem'' as the number of times an individual electron would need to be re-accelerated.

These requirement apply specifically to the upward-current regions above each footpoint, such that the energetic electrons propagate downward. As in a substorm, acceleration also occurs in the nearby downward-current regions where electrons are accelerated upward \citep{Eetal03}. 

\subsection{Return current and the number problem}

Th number problem  \citep{MB89} presents an important difference between flares and substorms. In both cases the power is $I\Phi$ with ${\dot N}=MI/e$ and $\varepsilon=e\Phi/M$, but with a large multiplicity ($M=10^6$ with the fiducial numbers assumed here) in flares and $M$ of order unity in substorms. In substorms, inverted-V electrons are accelerated in an auroral cavity \citep{BC79}, and as a result of ``charge starvation'' the flux tube contains essentially only the energetic electrons \citep{Eetal00}. In flares, the multiplicity is conventionally interpreted in terms of a return current replacing the electrons in the acceleration region $M$ times. An alternative interpretation is that the acceleration region and the hard X-ray source are co-located, and the same electrons are accelerated (and lose their energy emitting X-rays) a total of $M$ times \citep{Betal09}.

If the coaxial model developed above is driven hard enough, it can adjust to maximize the power released. If no return current is allowed, $b\ge0$, the limiting case is when all the current is confined to the inner region, which becomes as small as possible. This corresponds to $I_{\rm in}=I_0$, $I_{\rm out}=0$, $a\xi_1^2=1$, $b=0$. When a return current is allowed, $b<0$, an even larger power is possible. For $b<0$, $I_{\rm out}$ has the opposite sign to $I_{\rm in}$, and corresponds to a return-current for $\xi>\xi_1$, as illustrated in Fig.~\ref{fig:coaxial}. It follows from (\ref{Af7}) that the power increases with increasing $|b|$ for $b<0$. The model becomes untenable for $b$ large and negative. Nevertheless, this case gives an indication of how the number problem can be resolved. To show this, let us assume $b\approx-M$, and explore the implications.

According to (\ref{Af7}), the power in the Alfv\'enic flux increases $\propto(1-b)^2\approx M^2$ for $-b=M\gg1$. This can be attributed to the combination of two effects. First, the total current flowing in the region $r<\xi_1R$ is much greater than $I_0$, with (all but $I_0$ of) this direct current being provided by recycling the return current, which closes across field lines in the photosphere and in the energy-release region. Second, the large current $I_{\rm in}\approx MI_0$ implies a correspondingly large cross-field $\Phi\approx M\Phi_0$, which is implicit in the assumption of Alfv\'enic flux. With $M\approx10^6$, the extremely large values of $I$ and $\Phi$ are not tenable. In the coaxial model the direct current flows in a single current channel, at $r<\xi_1R$, and relaxing this assumption opens the possibility of formulating a model that can resolve the number problem.

Suppose that the direct current is of order $MI_0$ flowing in $M$ parallel channels, each carrying a current $I_0$. The return current flows between these channels, and is of magnitude $(M-1)I_0$. If the potential associated with each channel is $\Phi_0/M$, with $\Phi_0=\Sigma_AI_0$, the Alfv\'enic flux in each channel transports a power $I_0\Phi_0/M$, giving the total power $I_0\Phi_0$ after summing over the $M$ channels. In the acceleration region, this power is transferred to electrons with energy $\varepsilon=e\Phi_0/M$ at a rate $I_0/e$ in each channel, resulting in $MI_0/e$ electrons per second with this energy from the $M$ channels. The implied cross-field potential in any given channel is $\Phi_0/M\approx10^4\,$V, consistent with the observed energy, $\varepsilon$, of order $10\,$keV.  

Although these arguments are heuristic, they give a clear indication of how the number problem can be resolved. A detailed model incorporating these ideas is needed to make further progress.

\subsection{Development of $E_\parallel$}

An understanding of acceleration by $E_\parallel$ involves two complementary aspects: the macroscopic electrodynamics associated with the development of $E_\parallel$, and the microphysics associated with the specific (local and transient) structures with $E_\parallel\ne0$. Acceleration by $E_\parallel$ is widely accepted for auroral electrons in a substorm \citep{BF90}, where $E_\parallel$ can be observed in situ \citep{MK98}, and also for Io-related electrons in Jupiter's magnetosphere \citep{Setal03}. However, the microphysics involved in setting up and maintaining $E_\parallel$ is inadequately understood, with none of the many suggested mechanisms being entirely satisfactory \citep{B93}. As remarked in paper~1, in auroral acceleration $E_\parallel$ is highly structured on the microscale, with individual structures being modeled as, for example, phase-space holes  \citep{Eetal98b} and kinetic Alfv\'en waves \citep{Cetal03}. Discussion of acceleration by $E_\parallel$ in the solar corona also emphasizes microphysics \citep{BKB10}. The viewpoint adopted here is that important aspects of acceleration by $E_\parallel$ can be understood in terms of the macroscopic electrodynamics, independent of the microphysics that sustains $E_\parallel$. This view was expressed by \cite{H94}, who argued that the appeal to microphysics (double layers and anomalous resistivity) is required only to support other arguments that require the appearance of $E_\parallel$. 

One argument that an $E_\parallel$ must appear is that the downward Alfv\'enic flux, transporting the released magnetic energy, has a large ${\bi E}_\perp$ and associated vortex motion, ${\bi u}$,  and  these cannot be sustained through the denser chromospheric and photospheric plasma. As illustrated in Fig.~\ref{fig:auroral} for the auroral application, the surfaces of constant potential, corresponding to ${\bi E}_\perp$, must close above the conducting ionosphere or chromosphere, implying that $E_\parallel$ must develop. Another argument is that line-tying, associated with the  inertia of the denser regions of the ionosphere or chromosphere, precludes the twisting motion driven from above being maintained. The energy in the Alfv\'enic flux must be reflected, absorbed or converted into another form somewhere above the boundary where line-tying becomes effective. This implies that along a given field line, $u_\phi\ne0$ must reduce to zero, and this violates the frozen-in condition. The development of $E_\parallel$ is essential to allow a decoupling between two regions in relative motion perpendicular to ${\bi B}$, called fracturing  by \cite{H94,H07}. These arguments imply that there must be a region with $E_\parallel\ne0$ irrespective of the specific microphysics that sustains $E_\parallel$. In the model suggested above in connection with the return current and the number problem, the cross-field potential is $\Phi_0/M$, and the implication is that this leads to a parallel potential $\Phi_0/M$ along field lines in the acceleration region. The mocrophysics involved in how $E_\parallel$ develops is not directly relevant: the system must find some form of microphysics to satisfy the macroscopic requirements.

\section{Discussion and conclusions}
\label{section:conclusions}

In this paper, and in paper~1, a generic model for magnetic explosions is formulated, based on magnetospheric substorms, and applied to solar flares. Three ideas are central to the generic model: the separation into different scales (global, macro, micro) and into different  (energy-release, acceleration, current-closure) regions; the insistence that the intrinsic time dependence  be included explicitly (through ${\bf E}_{\rm ind}$, ${\bi u}_{\rm ind}$, ${\bf J}_{\rm disp}$, ${\bf J}_{\rm pol}$); and, that the electron acceleration is due to a parallel electric field, $E_\parallel$. 

What triggers a flare is an unsolved problem. The recognition that the inflow into the energy-release region is driven by the changing magnetic field offers a new way of thinking about the trigger. Prior to a flare the magnetic structure is in a steady state, with any pressure gradients balanced by the ${\bi J}\times{\bi B}$ force, with a very small angle between ${\bi J}$ and ${\bi B}$. Once a flare is triggered, the unbalanced Maxwell stress, mediated by ${\bf E}_{\rm ind}$ and ${\bf J}_{\rm disp}$,  drives the energy release, without requiring a mechanical force.  The trigger can be a perturbing pressure gradient that initiates inflow into the energy-release region, causing ${\bi B}$ to start to change with time. The changing ${\bi B}$ can then takes over, driving the magnetic energy into the energy-release region and powering the explosion.

It is assumed that the energy release occurs in two stages, involving two widely separated regions with an Alfv\'enic flux transporting the energy between them \citep{FH08}. The energy transport is modeled in terms of an Alfv\'enic energy flux associated in a current-carrying flux tube. A coaxial model is developed, in which the direct current becomes strongly concentrated near the axis of the cylinder. A limiting case of this model is when all the current is confined to narrow channel, and in this limit the total power released can be transported Alfv\'enically from the energy-release region to the acceleration region.

The ``number problem'' has presented a formidable challenge to all theories of electron acceleration in flares: the hard X-ray date imply either that the number of electrons accelerated is $M$ times the number available, or that the available electrons are re-accelerated $M$ times, with $M$ of order $10^6$. The observations required electrons with an energy $e\Phi_0/M$ accelerated at a rate ${\dot N}=MI_0/e$, such that the total power, $I_0\Phi_0$, is that released in the flare. Based on the coaxial model developed here, it is suggested that the number problem may be resolved by incorporating three changes. First, the single direct current channel is replacement by $M$ current channels, each carrying a current $I_0$. Second, there is a return current of magnitude $(M-1)I_0$ that flows between these channels, with the current lines closing in the photosphere and the energy-release region. Third, the cross-field potential associated with each channel is $\Phi_0/M$, and this potential develops along field lines in the acceleration region.

Another feature of the model is that there are both upward and downward current regions at each footpoint of the flaring flux tubes. The energetic electrons that produce hard X-rays as they precipitate are attributed to acceleration in the upward current region, and the electrons that escape to produce type~III radio bursts are attributed to acceleration in the downward current region. These electrons do not come from a single distribution and need not have the same properties. Energetic ions, accelerated downward in the downward current region, are necessarily slightly displaced from the electrons accelerated downward in the upward current region. Such effects need to be treated quantitatively in a detailed model.

\section*{Acknowledgments}

I thank Iver Cairns and Mike Wheatland for helpful comments.

\end{document}